\documentstyle[preprint,revtex]{aps}
\begin{document}
\begin{title}
Transverse response functions in the $\Delta$-resonance region
\end{title}
\author{E. Rost, C. E. Price and J. R. Shepard}
\begin{instit}
Department of Physics, University of Colorado, Boulder, CO 80309
\end{instit}
\begin{abstract}
We calculate transverse response functions for quasi-elastic
electron scattering at high momentum transfers in a relativistic
Hartree approximation in configuration space.  We treat the
excitation of the $\Delta$ resonance using its free mass and width.
Good agreement with experiment
is found in the dip region.
\end{abstract}
\newpage
\begin{narrowtext}
\section{INTRODUCTION}

In recent years several measurements\cite{barr}-\cite{chen} of separated
(e,e$'$) response functions, $S_L$ and $S_T$,
have been performed at high momentum transfer ($|{\bf q}|\ge$ 400 MeV/c).
The longitudinal response function, $S_L$, has been useful in studying
nuclear properties such as the Coulomb sum rule\cite{wal},
RPA correlations\cite{jrs} and vacuum fluctuation\cite{hpp}
effects.  Nuclear many-body theories agree with
the data at the $\sim$ 20\% level and the study of various effects
via comparison with experiment is meaningful.

On the other hand the measured transverse response functions, $S_T$, increase
at large excitation energies while nucleon-only theories
yield decreasing transverse response with energy.  It has long been suspected
that this effect is a result of $\Delta$ excitation and many calculations
have been performed extending the original work of Moniz\cite{mon}.
These calculations are typically performed nonrelativisically in momentum
space using harmonic oscillator wavefunctions for the expansion of
the $\Delta$-hole propagator\cite{kmo}.  A relativistic calculation has been
performed recently by Wehrberger
{\it et al.}\cite{wbb} who have calculated the electromagnetic excitation
of the $\Delta$ baryon in the framework of a quantum hadrodynamic model.
These calculations treat the $\gamma N \Delta$ coupling carefully using
a spin-3/2 $\Delta$ spinor.  However they employ a local density
approximation for the nuclear system along with a somewhat arbitrary
folding procedure to treat the $\pi N$ width of the $\Delta$.

In this paper we calculate the excitation of the $\Delta$ employing
a relativistic Hartree approximation (RHA) in configuration
space.  By using a non-spectral\cite{sr} description of the Green function
we are able to treat the continuum in our configuration space propagator
without
approximation or truncation.  While previous works have emphasized the
$\gamma N \Delta$ vertex at the expense of the nuclear structure part of the
calculation, we have chosen to treat the nuclear structure carefully and
to make some approximations in treating the vertex.   This permits us to
employ the successful RHA description of nuclear properties in
a straightforward way.

\section{FORMALISM---NUCLEON EXCITATION}

Following the treatment of Horowitz and Piekarewicz\cite{hp}
the inclusive electron-scattering cross section from an unpolarized
nuclear target is given by
\begin{equation}
 {{d^2\sigma}\over{d\Omega'dE'}} = \sigma_M \Bigl[
 {{Q^4}\over{{\bf q}^4}} S_L({\bf q},\omega) +
 \Bigl({{Q^2}\over{2 {\bf q}^2}}+\tan^2(\theta/2)\Bigr)
 S_T({\bf q},\omega) \Bigr],   \label{eq:aa}
\end{equation}
where $\sigma_M$ is the Mott cross section, $\omega = E - E'$ is the
electron energy loss, ${\bf q}$ is the three-momentum transfer,
$Q^2 = {\bf q}^2 - \omega^2$ is (minus) the space-like four-momentum
transfer, and $\theta$ is the scattering angle.  The nuclear response
to the weak electromagnetic probe is contained in the two structure
functions $S_L({\bf q},\omega)$ and $S_T({\bf q},\omega)$.  The
transverse response function in turn may be expressed in terms of the
polarization tensor using
\begin{equation}
 S_T({\bf q},\omega) = -{{1}\over{\pi}} [\delta_{ij}-{\bf{\hat q}}_i
{\bf{\hat q}}_j]
 {\rm Im} \Pi^{ij}({\bf q},{\bf q};\omega).  \label{eq:ab}
\end{equation}
In the Hartree approximation we obtain $\Pi$ by evaluating
\begin{equation}
 i\Pi^{\mu\nu}(x,y)=\langle\Phi_0|T[J^\mu(x) J^\nu(y)]|\Phi_0\rangle,
   \label{eq:ac}
\end{equation}
where $\Phi_0$ is the self-consistent Hartree ground state and
\begin{equation}
 J^\mu(q)=F_1(q^2) \gamma^\mu + F_2(q^2) {{i\kappa}\over{2 M}}
  \sigma^{\mu\nu} q_\nu,  \label{eq:ad}
\end{equation}
is the electromagnetic current.  The
evaluation of Eq.~\ref{eq:ac} is standard\cite{hp} and will not be repeated in
detail.
The result appropriate for the evaluation of structure functions is
\begin{eqnarray}
 \Pi^{\mu\nu}({\bf q},{\bf q};\omega) =&&\int d{\bf x}
 \,{\rm e}^{-i{\bf q\cdot x}}
 \int d{\bf y} \,{\rm e}^{i{\bf q\cdot y}}\nonumber\\
 &&\sum_h \Bigl[{\bar U}_h({\bf x}) J^\mu
 G_F({\bf x},{\bf y};\epsilon_h+\omega) J^\nu U_h({\bf y})\nonumber\\
 &&\phantom{\sum_h} + {\bar U}_h({\bf y}) J^\nu
 G_F({\bf y},{\bf x};\epsilon_h-\omega) J^\mu U_h({\bf x})\Bigr].
 \label{eq:ae}
\end{eqnarray}
where the sum is over hole states with energy $\epsilon_h$ and wavefunction
$U_h$.  The Feynman
propagator  $G_F$ in Eq.~\ref{eq:ae} is generated {\it non-spectrally}\cite{sr}
by solving the Dirac equation
\begin{equation}
  [\omega\gamma^0 + i{\bf \gamma\cdot\nabla} - M - \Sigma_H({\bf x})]
 G_F({\bf x},{\bf y};\omega) = \delta({\bf x}-{\bf y}), \label{eq:af}
\end{equation}
in the presence of the self-consistent Hartree field, $\Sigma_H$, and with
the appropriate boundary conditions\cite{sr,hp}.  The details of
calculating $\Sigma_H$ and $G_F$ in partial wave expansion are given
elsewhere\cite{hp}-\cite{srm}.

\section{FORMALISM---DELTA EXCITATION}

We now turn to generalizing the above treatment to include excitation of
the 3-3 nucleon resonance taken to be a $\Delta$ particle of mass
$M_\Delta$=1232 MeV/c$^2$ and decay width $\Gamma$=110 MeV.  The Feynman
propagator, $G_F$ in Eq.~\ref{eq:ae}, will then involve $\Delta$ states while
the hole states, $U_h$, will remain of nucleon character so that we
need to evaluate a $\gamma N\Delta$ transition amplitude written symbolically
as $\langle\Delta|J^\mu|N\rangle$.  This transition amplitude
may be obtained using the Rarita-Schwinger formalism\cite{js} for the
spin-3/2 $\Delta$ particle.  The conventional
choice yields\cite{wbb,js}
\begin{equation}
\langle\Delta|J^\mu|N\rangle = \Psi^{\Delta\beta}(k)
   \ F(q^2)\ \Gamma_\beta^\mu(q)\Psi(p),      \label{eq:ba}
\end{equation}
with
\begin{equation}
 \Gamma_\beta^\mu(q) = (-q_\beta g^{\mu\lambda}
 + q^\lambda g_\beta^\mu) M_\Delta\gamma_\lambda\gamma_5
 + (q_\beta k^\mu - q^\nu k_\nu g_\beta^\mu)\gamma_5. \label{eq:bb}
\end{equation}
and
\begin{eqnarray}
F(q^2)=&&{{-(M_\Delta+M_N)}\over{M_N({(M_\Delta+M_N)}^2-q^2)}}
\sqrt{3/2}\nonumber\\
&&\times 3.0 \Bigl( 1-{{q^2}\over{0.71 GeV^2}}\Bigr)^{-2}
   \Bigl( 1-{{q^2}\over{3.5 GeV^2}}\Bigr)^{-1/2}. \label{eq:bba}
\end{eqnarray}
Here $k$ is the outgoing 4-momentum of the $\Delta$, $p$ is the incoming
4-momentum of the nucleon, $q=k-p$, $\Psi$ is a nucleon spinor, and
$\Psi^{\Delta\beta}$ is a $\Delta$ spinor with an index $\beta$.

The operator in Eq.~\ref{eq:bb} is too complicated to be readily treated in
configuration space.  In order to reduce its complexity,
we note that the second term should be smaller than the first by roughly the
ratio of the momentum $k$ to $M_\Delta$. Also the first term involves
diagonal coupling in the Dirac spinors while the second term couples upper
and lower components (so again the first term should dominate).
We also note that for the cases considered here,
the space-like part of the 4-momentum, ${\bf q}$, is
larger than the time-like part $\omega$.
Hence we take
Eq.~\ref{eq:bb} to be of the form
\begin{equation}
 \Gamma_\beta^\mu(q) \longrightarrow  \alpha_0\  M_\Delta\
    |{\bf q}| \gamma_5 \gamma^\mu,  \label{eq:bbb}
\end{equation}
leading to
\begin{equation}
 \langle\Delta|J^\mu|N\rangle \simeq
\alpha(|{\bf q}|)\ \Psi^\Delta(k)\,\gamma_5 \gamma^\mu\,\Psi(p), \label{eq:bc}
\end{equation}
with
\begin{equation}
\alpha(|{\bf q}|)\equiv\alpha_0\ F(q^2)\ M_\Delta\ |{\bf q}|. \label{eq:bca}
\end{equation}
This form preserves the $\pi N$ character of the $\Delta$ with its
$\gamma_5$ term and preserves the 4-vector character of the amplitude
(via the $\gamma^\mu$)
but the $\Delta$ is now effectively treated as a spin-1/2 spinor.
A constant factor $\alpha_0$ has been inserted since the simplification
in Eq.~\ref{eq:bbb} is severe.  Surprisingly, we find that $\alpha_0 = 1.0$
is adequate
for all cases that we studied.
We also find that the value of
$\alpha(|{\bf q}|)$ varies slowly for the
cases considered in this work (e.g., for the $^{12}$C dip region it
varies from $\sim 0.60$ to $\sim 0.75$.)

Another modification involves the
propagator which, for Delta excitation, is generalized to read
\begin{equation}
 [(\omega+i\Gamma/2)\gamma^0 + i{\bf \gamma\cdot\nabla} - M_\Delta
- \Sigma_H({\bf x})]
 G_F({\bf x},{\bf y};\omega) = \delta({\bf x}-{\bf y}), \label{eq:bd}
\end{equation}
{\it i.e.}, an imaginary term is added to the energy and
the $\Delta$ mass is used.
The $i\Gamma/2$ addition to the energy in Eq.~\ref{eq:bd} is
significant in that it leads to imaginary terms in $\Pi$
and hence non-vanishing $S_T$ below the $\Delta$ threshhold.  The
second term in Eq.~\ref{eq:ae} involves $\Delta$ baryons in the nuclear
ground state.  We have chosen to omit this term---fortunately
its contribution using various prescriptions is, at most,
a few percent.
The Hartree field, $\Sigma_H$, is taken to be
the same as that obtained self-consistently for the nucleons.  The
effects of this choice as well as the other calculational details
will be discussed in the following section.

\section{STUDY OF CALCULATIONAL SENSITIVITIES}

In Fig.~1 we show the transverse response for $^{12}$C as a function
of energy loss at a three-momentum transfer $|{\bf q}|$=550 MeV/c calculated
using the mean-field parameters of ref.~\cite{hp} ($g^2_s$=109.626, $m_s$=520,
$g^2_v$=190.431, $m_v$=783.)  The dotted line
shows the calculated response involving nucleon excitation alone and is
identical to the corresponding curve in ref.~\cite{hp}.  The dashed curve
presents
the calculation using {\it only} $\Delta$ excitation as outlined above and
the solid curve is the incoherent sum of the two.  The agreement with the data
is suprisingly good and may be fortuitous.  Indeed we expected that the
severe approximation involved in the simplification of Eq. \ref{eq:bbb} might
require an effective normalization constant, $\alpha_0$,
for which we have found the value of unity to be adequate.

We note that the Delta is propagated in our model (see Eq. \ref{eq:bd}) with
its
{\it free} width $\Gamma$.  This ignores medium effects of $\Delta$
propagation such as Pauli blocking and pion absorption.  Such effects
have been calculated (see, e.g., ref.~\cite{ko})  in a momentum-space
$\Delta$-hole formalism.  A subsequent paper\cite{cl} observed that
most of these effects
could be roughly accounted for by a simple shift ($-30 - i 40$) MeV in
the position of the Delta (the $-30$ MeV real shift corresponds to a
simplified self-energy $\Sigma$ in our notation.)

In Fig.~2 we show the effect of various alternate prescriptions for the
the Delta parameters in Eq.~\ref{eq:bd} as applied to the $\Delta$-excitation
part of Fig.~1.  The dot-dashed line was calculated using a
Delta width reduced by $40$ MeV as suggested in Ref.~\cite{cl}.  Finally
the dotted curve is calculated with the
extreme assumption of setting $\Sigma_H$ to $0$.  It is interesting that the
difference between the dashed and dotted curves is about equal to a constant
$-30$ MeV shift as suggested by Chen and Lee\cite{cl}.

The significance of the curves in Fig.~2 is that they have similar shapes
and thus would all give comparable results in the dip region with a
small modification of $\alpha_0$.  The falloff
in the dip region, say from $300$ MeV to $200$ MeV, is about a factor~$4$.
The Gaussian folding prescription of Ref.~\cite{wbb} leads to a falloff
factor of about $8$ over the same range and the results of
$\Delta$-hole momentum space calculations\cite{ko}
also produce a Delta ``line shape''
which falls off sharply with energy below threshold.  This different rate
of falloff is the reason that our calculations are able to reproduce
the experimental data in the dip region.

Although we are not able at present to improve on the vertex simplification in
Eq.~\ref{eq:bbb}, we can make some assessment of its importance by performing
the
calculations with an even simpler prescription obtained by removing
all lower components from the calculations so that the
vertex is essentially the operator ${\bf \sigma}$ taken between
non-relativistic wavefunctions.  The effect of this truncation
is a very minor ($\sim$ $1$ to $2$ \%) change in $S_T$ which would
be indistinguishable from the dashed curve in Fig.~2.  Our relativistic
formulation is used mainly for its convenience in computing nuclear
structure realistically with a minimal number of parameters.
{ \it We believe that our results for $S_T$ in the dip region are dominated
by our employment of a configuration
space calculation which includes a $\Delta$ decay width and which
treats the nuclear radius, surface and binding energies accurately.}

\section{RESULTS AND CONCLUSIONS}

Figures~3--5  present RHA calculations at $|{\bf q}|$=400 MeV/c for $^{12}$C,
and at $|{\bf q}|$=550 MeV/c for $^{40}$Ca and $^{56}$Fe (treated approximately
by assuming $^{56}$Fe is equivalent to a closed-shell $^{56}$Ni nucleus.)
The agreement with data in the Delta-excitation and the dip region is
surprisingly good considering that all employ the $\alpha_0=1$
normalization of the simplified $\gamma N \delta$ vertex.

Our calculations have not treated multinucleon interaction terms such as
meson exchange currents and our results do not preclude effects such as
those calculated by Dekker, Brussaard and Tjon\cite{dbt}.  However
such calculations should treat the single nucleon and single Delta excitation
realistically if they are to be compared with experimental data.

\newpage
\figure{
Transverse response functions for $^{12}$C at $|{\bf q}|$=550 MeV/c. The
dotted curve is the RHA result using only nucleon excitation.  The dashed
curve is the RHA result with only $\Delta$ excitation as described in the
text.  The solid curve is the sum of the dotted and dashed curves.}

\figure{
Transverse response function for $^{12}$C at $|{\bf q}|$=550 MeV/c with only
$\Delta$ excitation. The dashed curve represents our ``standard'' RHA
calculation as in Fig~1.  The dot-dashed curve is calculated with a shift
of the $\Delta$ width by $-40$ MeV.  The
dotted curve is calculated using a zero self-energy $\Sigma$.}

\figure{
Transverse response functions for $^{12}$C at $|{\bf q}|$=400 MeV/c. The curves
are described in the caption to Fig.~1.}

\figure{
Transverse response functions for $^{40}$Ca at $|{\bf q}|$=550 MeV/c. The
curves
are described in the caption to Fig.~1.}

\figure{
Transverse response functions for $^{56}$Fe at $|{\bf q}|$=550 MeV/c. The
curves
are described in the caption to Fig.~1.}

\end{document}